\documentclass[12pt,twoside]{article}
\usepackage{fleqn,espcrc1}

\usepackage{graphicx}
\newcommand{\AmS}{{\protect\the\textfont2
  A\kern-.1667em\lower.5ex\hbox{M}\kern-.125emS}}

\title{R-Process Nucleosynthesis in Core-Collapse Supernova Explosion}

\author{
T. Kajino \address{National Astronomical Observatory and
Graduate University for Advanced Studies, 2-21-1 Osawa, Mitaka, Tokyo 181-8588, Japan}
\address{University of Tokyo, Graduate School of Science, Department of Astronomy, \\
7-3-1 Hongo, Bunkyo-ku, Tokyo 113-0033, Japan},
~S. Wanajo \address{Sophia University, Faculty of Science and Technology, Department of Physics, \\
7-1 Kioi-cho, Chiyoda-ku, Tokyo 102-8554, Japan},
~G. J. Mathews \address{University of Notre Dame, Department of Physics and Center for Astrophysics, \\
Notre Dame, IN46556, U.S.A.}
}

\begin{document}

\maketitle
\begin{abstract}

We study the r-process nucleosynthesis in neutrino-driven winds of gravitational core collapse SNeII.  Appropriate physical conditions are found for successful r-process nucleosynthesis, 
which meet with several features of heavy elements discovered recently in metal-deficient halo stars.  
We find also several difficulties which are not explained in the present wind models.  We discuss quests for new insights in nuclear physics, astrophysics, and astronomy.

\end{abstract}

\section{Introduction}

Stars with various masses provide a variety of production sites 
for intermediate-to-heavy mass elements.  Very massive stars  
$\geq 10 M_{\odot}$ culminate their evolution by supernova (SN) explosions 
which are also presumed to be most viable candidate for the still unknown 
astrophysical site of r-process nucleosynthesis.  Even in the 
nucleosynthesis of heavy elements, initial entropy and density 
at the surface of proto-neutron stars are so high 
that nuclear statistical equilibrium (NSE) favors production of abundant light nuclei.  
In such explosive circumstances of the so called hot-bubble scenario,
not only heavy neutron-rich nuclei but also light unstable
nuclei play a significant role.

Studies of the origin of the r-process elements are critical in cosmology, too. 
There still remains potentially a serious problem that the cosmic age 
of the expanding Universe derived from cosmological parameters may be shorter 
than the age of the oldest globular clusters.  Since both age estimates
are subject to the uncertain cosmological distance scale, 
an independent method has long been needed.  Thorium (Th) and Uranium (U), 
which are typical r-process elements and have half-lives of 14 Gyr (Th) and
4.5Gyr ($^{238}$U), have recently been detected 
along with other elements in very metal-deficient stars~\cite{sneden96,cayrel00}.
If we model the r-process nucleosynthesis in these first-generation
stars, these long lived radioactive nuclei can be used as good cosmochronometers
completely independent of the uncertain cosmological distance scale.
  
We discuss in this article that supernova 
explosions of very massive stars could be a viable site for r-process 
nucleosynthesis.  In sect. 2, we explain how to model the neutrino-driven winds.
The details of explosive nucelsynthesis in the hot-bubble is discussed in sect. 3.
Although our theoretical study has confirmed a successful r-process
in SNeII, there are several findings which are still difficult to explain.
We discuss several quests for nuclear physics and astrophysics in sect. 4, 
and astronomy in sect. 5.

\section{Neutrino-Driven Winds in Type-II Supernovae}

Recent measurements using high-dispersion spectrographs with large 
Telescopes or the Hubble Space Telescope have made it possible
to detect minute amounts of heavy elements in faint metal-deficient 
([Fe/H] $\le$ -2) stars~\cite{sneden96}. The discovery of r-process elements in these 
stars has shown that the relative abundance pattern for the mass region 
120 $\le$ A is surprisingly similar to the solar system r-process abundance 
independent of the metallicity of the star.
Here metallicity is defined by 
[Fe/H] = log[N(Fe)/N(H)] - log[N(Fe)/N(H)]$_{\odot}$. It obeys the
approximate relation t/10$^{10}$yr $\sim$ 10$^{[Fe/H]}$. 
The observed similarity strongly suggests that the r-process occurs 
in a single environment which is independent of progenitor metallicity.
Massive stars with 10$M_{\odot} \le M$ have a short life, 
$\sim 10^7$ yr, and eventually end up as violent supernova explosions, 
ejecting material into the interstellar medium early on in the
history of the Galaxy.   
However, the iron shell in SNe is excluded from being the 
r-process site because of the observed metallicity independence.  

Hot neutron stars just born in the gravitational core collapse 
SNeII release most of their energy as neutrinos during the Kelvin-Helmholtz 
cooling phase.  An intense flux of neutrinos heat the material near the
neutron  
star surface and drive matter outflow (neutrino-driven winds).
The entropy in these winds is so high that the NSE favors a plasma
which consists of mainly free nucleons and alpha particles rather than
composite nuclei like iron.  
The equilibrium lepton fraction, 
$Y_e$, is determined by a delicate balance between 
$\nu_e + n \rightarrow p + e^-$ and $\bar{\nu}_e + p \rightarrow n + e^+$,
which overcomes the difference of chemical potential between $n$ and $p$,
to reach $Y_e \sim$ 0.45.  R-process nucleosynthesis occurs because there are
plenty of free neutrons at high temperature. 
This is possible only if seed elements are produced in the correct 
neutron-to-seed ratio before and during the r-process.

Although Woosley et al.~\cite{woosley94} demonstrated a profound possibility 
that the r-process could occur in these winds, several difficulties were 
subsequently identified.  First, independent non-relativistic numerical
supernova models~\cite{witti94} have difficulty producing
the required entropy in the bubble, S/k $\sim$ 400.  
Relativistic effects may not be enough to increase 
the entropy dramatically~\cite{qian96,cardall97,otsuki00}.  
Second, even should the entropy be high enough, the effects of neutrino 
absorption $\nu_e + n \rightarrow p + e^-$ and 
$\nu_e + A(Z,N) \rightarrow A(Z+1,N-1) + e^-$ 
may decrease the neutron fraction during the nucleosynthesis process.
As a result, a deficiency of free neutrons could prohibit the
r-process~\cite{meyer95}.   
Third, if neutrinos are massive and have appropriate mixing parameters, 
energetic $\nu_{\mu}$ and $\nu_{\tau}$ are converted into $\nu_e$ due to 
flavor mixing. This activates the $\nu_e + n \rightarrow p + e^-$ process 
and results in a deficiency of free neutrons.

In order to resolve these difficulties, we have studied~\cite{otsuki00,wanajo01}
neutrino-driven winds in a Schwarzschild geometry under the  
reasonable assumption of spherical steady-state flow.
\newpage
\begin{figure}[htb]
\hspace{2.5cm}\includegraphics[width=110mm]{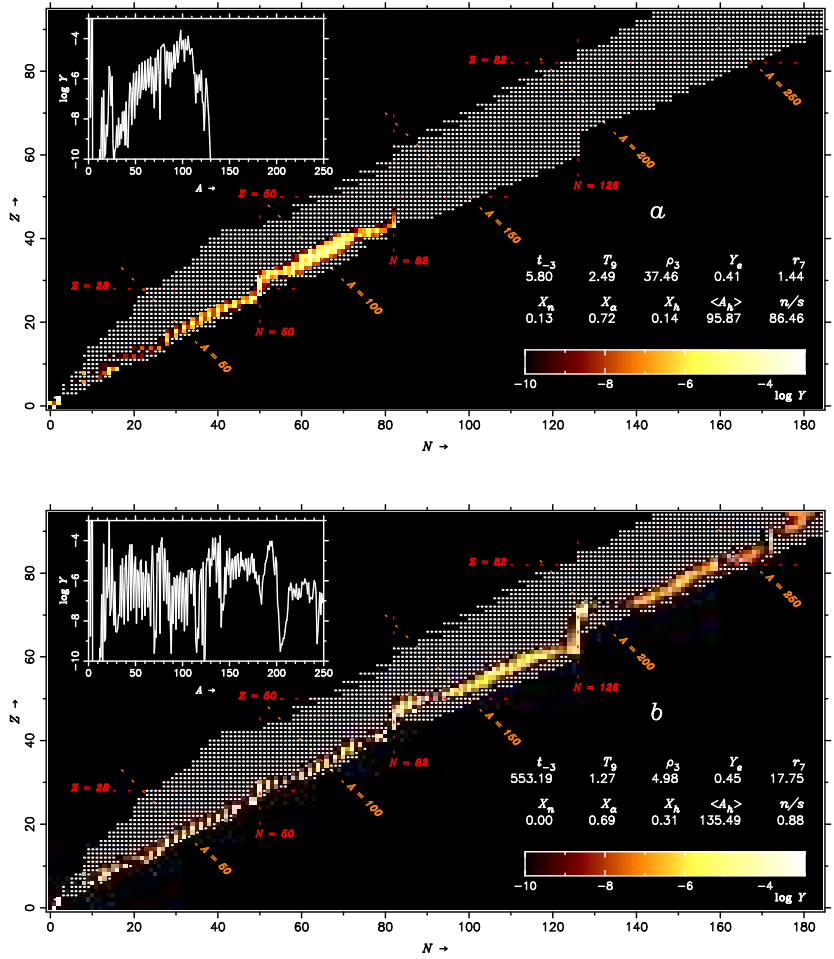}
\caption{Nuclear reaction flow patterns in Z-N plane, and abundances 
in the insets at the time;
(a) t $\approx 20$ ms during the $\alpha$-process (upper), 
and (b) t $\approx 567$ ms during the r-process (lower).  
Time zero refers to the time of core bounce.  
Highlight abundance scales to log$Y$, where $Y$ is the number fraction
of each nucleus.  
White dots show the nuclei included in the reaction network 
code~\cite{otsuki00,wanajo01}.
The neutrino-driven wind model used is for
$L_{\nu} = 10^{52}$ ergs/s and $M = 2 M_{\odot}$~\cite{otsuki00}.
\label{flow}}
\end{figure}
\noindent
The parameters in the wind models are the mass of neutron star, $M$,
and the neutrino luminosity, $L_{\nu}$.  
The entropy per baryon, S/k, in the asymptotic regime
and the expansion 
dynamic time scale, $\tau_{dyn}$, which is defined
as the duration time of the $\alpha$-process when the temprature drops from
T $\approx$ 0.5 MeV to 0.5/e MeV,
are calculated from the solution of the hydrodynamic equations.
Then, we carried out r-process nucleosynthesis calculations in our wind model.
We found~\cite{otsuki00} that the general relativistic effects make $\tau_{dyn}$ 
much shorter, although the entropy increases by about 40 \% from the 
Newtonian value of S/k $\sim$ 90.
By simulating many supernova explosions, we have found some interesting conditions  
which lead to successful r-process nucleosynthesis,
as to be discussed in the following sections. 
\begin{center}
\begin{figure}
\hspace{2.5cm}\includegraphics[width=75mm,angle=-90]{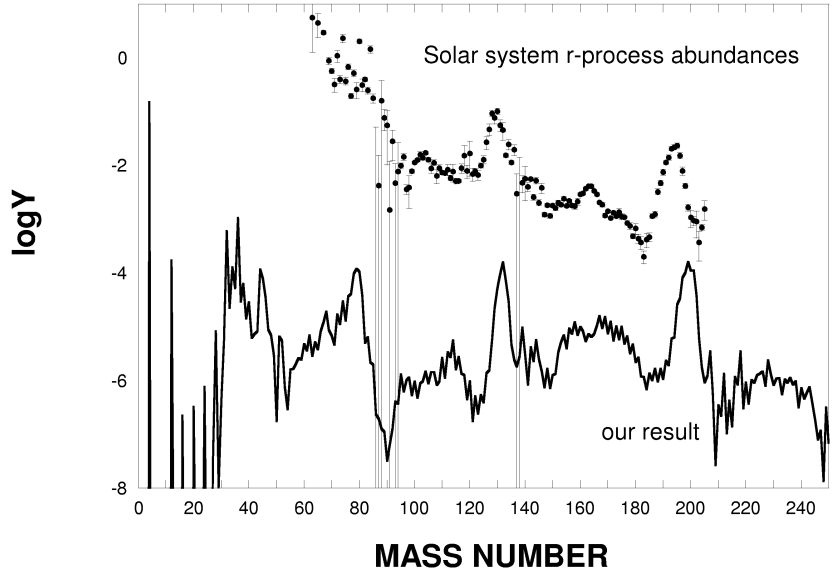}
\caption{
R-process abundance~\cite{otsuki00} (solid line) as a function of atomic mass number
$A$ compared with the solar system r-process abundance (filled circles) 
from K\"appeler, Beer, \& Wisshak~\cite{kappeler89}.
The neutrino-driven wind model used is for
$L_{\nu} = 10^{52}$ ergs/s and $M = 2 M_{\odot}$.
The solar system r-process abundance is shown in arbitrary unit. 
\label{abundance}}
\end{figure}
\end{center}

\section{Nucleosynthesis}

\subsection{R-process and the key reaction}

Previous r-process calculations~\cite{woosley94,meyer92} had shortcoming that the 
seed abundance distribution was 
first calculated by using
smaller network for light-to-intermediate mass elements,
and then the result was connected further to another r-process network
in a different set of the computing run.
For this reasaon it was less transparent to interpret 
the whole nucleosynthesis process.
This inconvenience happened because it was numerically too heavy to run
both $\alpha$-process and r-process in a single network code for too huge
number of reaction couplings among $\sim 3000$ isotopes.
Our nucleosynthesis calculation is completely free from this complexity
because we exploited a fully implicit single network code which 
is applied to a sequence of the whole processes of NSE - $\alpha$-process - r-process.
Shown by white dots in Fig.~1 are the isotopes included 
in our reaction network~\cite{otsuki00,wanajo01}.

Let us remind the readers that there were at least three difficulties 
in the previous theoretical studies of the r-process.  
The first difficulty is that an ideal, high entropy 
in the bubble S/k $\sim$ 400~\cite{woosley94} was hard to be achieved in 
other simulations~\cite{witti94,qian96,cardall97,otsuki00}.

The key to resolve this difficulty is found with the short dynamic time scale  
$\tau_{dyn}\sim$ 10 ms in our models of the neutrino-driven winds.  
As the initial nuclear 
composition of the relativistic plasma consists of neutrons and protons, 
the $\alpha$-burning begins when the plasma temperature cools below 
T $\sim$ 0.5 MeV.  The $^4$He$(\alpha \alpha,\gamma)^{12}$C reaction 
is too slow at this temperature, and an alternative nuclear reaction path 
$^4$He$(\alpha n,\gamma)^9$Be$(\alpha,n)^{12}$C triggers explosive 
$\alpha$-burning to produce seed elements with A $\sim$ 100 (Fig.~1a).  
Therefore, the time scale for nuclear reactions is regulated by 
the $^4$He$(\alpha n,\gamma)^9$Be reaction. It is given by 
$\tau_N \equiv \left(\rho_b^2 Y_{\alpha}^2 Y_n \lambda(\alpha \alpha n
\rightarrow ^9{\rm Be}) \right)^{-1}$.
If the neutrino-driven winds fulfill the condition 
$\tau_{dyn} < \tau_N$, then fewer seed nuclei are produced 
during the $\alpha$-process with plenty of free neutrons left over when
the r-process begins at T $\sim$ 0.2 MeV.  The high neutron-to-seed  
ratio, $n/s \sim 100$, leads to appreciable production of r-process 
elements, even for low entropy S/k $\sim$ 130, producing both the 2nd $(A \sim
130)$ and 3rd $(A \sim 195)$ abundance peaks and the hill of rare-earth elements 
$(A \sim 165)$ (Fig.~1b and Fig.~2).
\begin{figure}[htb]
\hspace{3.0cm}\includegraphics[width=100mm]{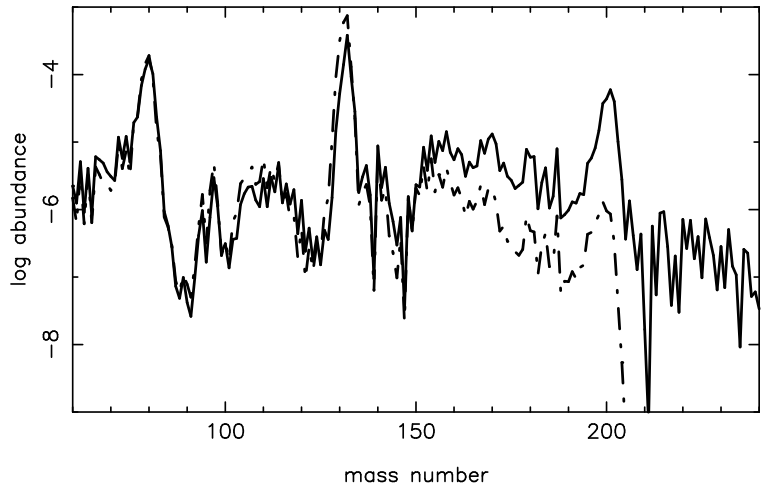}
\caption{
The same as in Fig. 2, but for the neutrino-driven wind model of
$L_{\nu} = 5 \times10^{52}$ ergs/s.
Solid line represents the result by using the Woosley \& Hoffman
rate~\cite{woosley92} of the  $^4$He$(\alpha n,\gamma)^9$Be reaction, 
and long-dashed line shows the result for this rate multiplied by factor 2,
as suggested by the recent experiment of 
Utsunomiya et al.~\cite{utsunomiya01}.
\label{abundance2}}
\end{figure}

The three body nuclear reaction cross section for $^4$He$(\alpha n,\gamma)^9$Be 
is one of the poorly determined nuclear data which may 
alter the r-process nucleosynthesis yields.  
The inverse process has recently been studied experimentally 
by Utsunomiya et al.~\cite{utsunomiya01}, 
and photodisintegration cross section of $^9$Be
has been measured with better precision than those of the previous experiments.
Applying the principle of the detailed balance to this process, one can estimate the 
cross section for $^4$He$(\alpha n,\gamma)^9$Be.
They found that the thermonuclear reaction rate is almost twice as big as that of 
Woosley and Hoffman~\cite{woosley92} but in resonable agreement with the recent
compilation of Angulo et al.~\cite{angulo99}.
However, there still remain several questions on the consistency between 
their result and electron-scattering experiments, on the contribution from the narrow 
resonance $J^{\pi} = 5/2^{-}$ (2.429 MeV), etc.
It is also a theoretical challenge to understand the reaction mechanism and 
the resonance structure because two different channels, 
$^8$Be + n and $^5$He + $\alpha$, contribute to this process.

Therefore, we show two calculated results in Fig. 3:
The solid line displays the result obtained by using the Woosley and Hoffman 
cross section~\cite{woosley92}, assuming a $^8$Be + n structure
for $^9$Be.  We also calculated the r-process by multiplying
this cross section by factor of 2 (long-dashed line).  This makes a drastic change in
the r-process yields in the 3rd $(A \sim 195)$ abundance peak.
More theoretical and experimental studies of the 
$^4$He$(\alpha n, \gamma)^9$Be reaction are highly desired. 

\subsection{Neutrino-nucleus interactions}

Neutrino interactions with nucleons and nuclei take the key to resolve 
the second difficulty which was pointed out in sect. 1.  The difficulty is that
the effects of neutrino absorptions $\nu_e + n \rightarrow p + e^-$ and 
$\nu_e + A(Z,N) \rightarrow A(Z+1,N-1) + e^-$ 
during the $\alpha$-process
may induce a deficiency of free neutrons 
and break down the r-process conditions~\cite{meyer95}. 
These two types of neutrino interactions control most sensitively 
the electron fraction and the neutron fraction
in a neutron-rich environment.   
In order to resolve this difficulty, we have 
updated the electron-type neutrino capture
rates for all nuclei and electron-type anti-neutrino capture
rate for free protons~\cite{qian97,meyer98}.   

The new r-process calculation proves to be almost invariant.  
One can understand this robustness of the succesful r-process 
in the following way:
The specific collision time for neutrino-nucleus interactions is given by
\begin{equation}
\tau_{\nu} \approx 201 \times L_{\nu ,51}^{-1} \times
\left(\frac{\epsilon_{\nu}}{\rm MeV}\right) 
\left(\frac{r}{\rm 100km}\right)^2 
\left(\frac{\langle\sigma_{\nu}\rangle}{\rm 10^{-41}cm^2} \right)^{-1} ms,
\label{eqn:tau}
\end{equation}
where $L_{i,51}$ is the individual neutrino or
antineutrino luminosity in units of $10^{51}$ ergs/s, 
$\epsilon_i=<E_i^2>/<E_i>$ in MeV
$(i=\nu_e,~~\bar{\nu}_e,~~etc. )$, 
and $\langle \sigma _{\nu} \rangle$ is the cross
section averaged over the neutrino energy spectrum. 
At the $\alpha$-burning site of r $\approx$ 100 km
for $L_{\nu ,51} \approx 10$, $\epsilon_{\nu_e}=12~\rm{MeV}$,
and $\langle \sigma _{\nu} \rangle \approx 10^{-41} cm^2$, 
$\tau_{\nu_e}$(r=100 km) turns out to be $\approx 240$ ms.  
This collision time is larger than the expansion dynamic time scale;
$\tau_{dyn} \approx$ 10 ms $\ll$ $\tau_{\nu_e}$(r=100 km) $\approx 240$ ms.
Because there is not enough time for $\nu_e$'s 
to interact with n's in such rapidly expanding neutrino-driven wind, 
the neutron fraction is insensitive to the neutrino absorptions.
(Note, however, that the opposite condition 
$\tau_{dyn} > \tau_{\nu_e}$ could apply to 
slowly expanding winds as adopted 
in the previous studies~\cite{woosley94,witti94,meyer92}.
In this case the r-process is strongly affected by the
neutrino interactions~\cite{meyer95}. )
 
One might wonder if our dynamic time scale $\sim 10$ ms 
is too short for the wind to be heated by neutrinos.
Careful comparison between proper expansion time and specific
collision time for the neutrino heating is needed in order to
answer this question.
Otsuki et al.~\cite{otsuki00} 
have found that the supernova neutrinos transfer their
kinetic energy to the wind most effectively just above the neutron star
surface at $10 {\rm km} \leq r < 20 {\rm km}$. 
Therefore, one should refer here to the duration
time for the wind to reach the $\alpha$-burning site, $\tau_{\rm heat}$,
rather than to $\tau_{\rm dyn}$. 
One can estimate this heating time scale 
\begin{equation}
\tau_{\rm heat} =
\int^{r_f}_{r_i}\frac{dr}{u},
\label{eqn:timeheat}
\end{equation}
where u is the fluid velocity of the wind.  By setting the radius of 
the neutron star surface to $r_i=10$ km and $r_f=100 $ km, 
we get $\tau_{\rm heat} \approx$ 30 ms.
The collision time $\tau_{\nu}$ is
given by Eq.~(\ref{eqn:tau}) by setting $L_{\nu ,51} \approx 10$, 
$\epsilon _{\nu}=(\epsilon _{\nu_ e} + \epsilon _{\bar{\nu}_e} )/2
= (12 + 22)/2 = 17$ MeV,
r $\approx$10 km, and $\langle \sigma _{\nu} \rangle \approx 10^{-41} cm^2$.
Let us compare $\tau_{\rm heat}$ and $\tau_{\nu}$ to one another:
\begin{equation}
\tau_{\nu} \approx 3.4 {\rm ms} ~\ll~ \tau_{\rm heat} \approx 30 {\rm ms}.
\label{eqn:taucomp}
\end{equation}
We can thus conclude that there is enough time for the expanding wind to be
heated by neutrinos even with short dynamic time scales
for the $\alpha$-process, $\tau_{\rm dyn} \sim 10$ ms.

\subsection{Neutrino oscillation}

The third difficulty discussed in sect. 1 is that 
the energetic massive $\nu_{\mu}$ and $\nu_{\tau}$
may change their flavors to emerge as $\nu_e$ due to the MSW effect.  
Although the luminosity of each type of neutrino $L_i
~~(i=\nu_e,~\bar{\nu}_e,~\nu_{\mu},~\bar{\nu}_{\mu},~\nu_{\tau},~\bar{\nu}_{\tau} 
)$ is similar,
average energies are different from one another;
$\epsilon_{\nu_e}=12~\rm{MeV}$,
$\epsilon_{\bar{\nu_e}}=22~\rm{MeV}$, and
$\epsilon_{\nu}=\epsilon_{\bar{\nu}}=34~\rm{MeV}$ 
for the other flavors~\cite{qian96}.  
Because of this difference, if the flavor mixing happened, 
newly converted $\nu_e$ were more energetic so that
the $\nu_e + n \rightarrow p + e^-$ reaction would decrease
the neutron fraction. 

We have recently shown that the neutrino flavor oscillation destroys the 
r-process condition only if the mixing parameters satisfy 
0.3 eV$^2 \le \Delta m^2$~\cite{otsuki01}.  
Although recent experiments of the atmospheric neutrinos and the 
missing solar neutrinos have indicated much smaller $\Delta m^2$, 
the LSND experiment suggests remarkably larger $\Delta m^2$
which overlaps with our interesting parameter region.  
We should wait for more experiments.

\subsection{Extended reaction network}

The r-process is thought to proceed after the pile up of seed nuclei 
produced in the $\alpha$-process at higher temperatures
$T_9 \approx 5 \sim 2.5$.
Since charged-particle reactions, which reassemble nucleons into $\alpha$-particles 
and $\alpha$-particles into heavier nuclei (i.e. $\alpha$-process),
are faster than the neutron-capture flow which is regulated by beta-decays,
the light-mass neutron-rich nuclei were presumed to be unimportant.

However, we have recently found~\cite{kajino00,terasawa01} that even light neutron-rich
nuclei progressively play the significant roles in the production of seed nuclei.
The nuclear reaction network used in the previous studies~\cite{woosley94,meyer92} 
included only limited number of light unstable nuclei, $^{3}$H, $^{7}$Be, $^{8,9}$B,
$^{11,14}$C, $^{13}$N, $^{15}$O, $^{18,20}$F, $^{23,24}$Ne, and so on.
We therefore need to extend the network code so that it covers 
all radioactive nuclei to the neutron-drip line.
We take the rates of charged particle reactions from those used in the
Big-Bang nucleosynthesis calculations~\cite{boyd89,kajino90a,kajino90b,orito97}
and the NACRE compilation~\cite{angulo99}.

Let us briefly discuss the results of the r-process calculations,
using the extended reaction network~\cite{terasawa01}.
At early epoch of the wind expansion, t $\le$ a few dozens ms, 
both temperature and density are so high that the charged particles
interact with one another to proceed nucleosynthesis around the $\beta$-stability line 
in the light-mass region A $\le$ 20.
There are plenty of protons and $\alpha$-particles as well as neutrons
at this epoch, and the main reaction flow is triggered by
$^4\rm{He}(\alpha n,\gamma)^9\rm{Be}$~\cite{woosley94,woosley92}:
\begin{eqnarray}
&^4\rm{He}(\alpha n,\gamma)^9\rm{Be}(\alpha,n)^{12}\rm{C}(n,\gamma)^{13}\rm{C}(\alpha,n)^{16}\rm{O}(n,\gamma)^{17}\rm{O}(\alpha,n)^{20}\rm{Ne}~~or~~^{16}\rm{O}(\alpha,\gamma)^{20}\rm{Ne} ...
\end{eqnarray}
\begin{eqnarray}
&^4\rm{He}(\alpha n,\gamma)^9\rm{Be}(\alpha,n)^{12}\rm{C}(n,\gamma)^{13}\rm{C}(n,\gamma)^{14}\rm{C}(n,\gamma)^{15}\rm{C} ... 
\end{eqnarray}
However, at relatively later epoch when the temperature drops below T = 0.5/e MeV
even after the $\alpha$-rich freeze out, 
a new reaction path~\cite{kajino00,terasawa01} starts operating:
\begin{eqnarray}
&^3\rm{H}(\alpha,\gamma)^7\rm{Li}(n,\gamma)^8\rm{Li}(\alpha,n)^{11}\rm{B}(n,\gamma)^{12}\rm{B}(n,\gamma)^{13}\rm{B}(n,\gamma)^{14}\rm{B}(n,\gamma)^{15}\rm{B}(e^- \nu)^{15}\rm{C} ...
\end{eqnarray}
This new reaction flow path also takes appreciable flux of baryon number 
and continuously produces seed nuclei.  The classical r-process like flow, (n,$\gamma$) 
followed by beta decay, has already started from light nuclei.
This is a very different result from the previous picture that
the r-process starts from only intermediate-mass seed nuclei $A \approx 100$.

Since we do not have much information of $(2n,\gamma)$ reactions,
we assumed in this paper that the (n,$\gamma$) and ($\gamma$,n) reactions
are in thermal equilibrium for the first neutron and then the second
neutron is subsequently captured.  Although inclusion of the
(2n,$\gamma$) reactions did not change drastically the final result,
these rates are presumed to be lower limit.  More elaborated theoretical
calculation of the (2n,$\gamma$) reaction has suggested that the
di-neutron correlation may increase these cross sections.
In the present calculations the yields of even the most neutron-rich isotopes 
were found to be abundant~\cite{terasawa01}.
Photodisintegration reactions of $^6$He and $^8$He and their
electromagnetic structure also have been studied 
experimentally~\cite{aumann99,iwata00}.  
Extensive measurements of the nuclear
properties of neutron-rich nuclei near the drip line, including
$^{17,19}$B and $^{29,31}$F, are yet to be carried out.

There are several branching points between (n,$\gamma$) and
$(\alpha$,n) reactions.  They are at $^{18}\rm{C}$,
$^{24}\rm{O}$, $^{36}\rm{Mg}$, etc.  
Experimetal studies to measure these reaction cross sections are highly desirable. 
%
%
\begin{center}
\begin{figure}
\hspace{0.5cm}\includegraphics[width=150mm,angle=0]{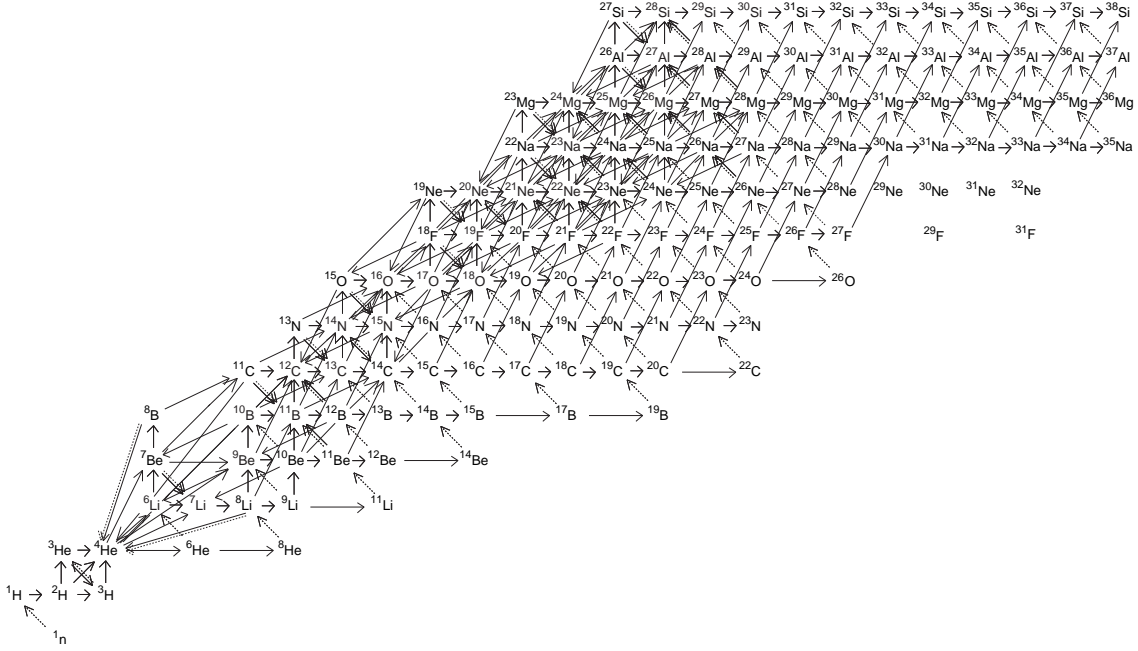}
\caption{Extended nuclear reaction network for explosive 
nucleosynthesis~\cite{terasawa01}.
For $\alpha$-process conditions at earlier times the main reaction flows are
triggered by the $^4\rm{He}(\alpha n,\gamma)^9\rm{Be}$ reaction
and two flow paths (4) and (5) follow.
For r-process conditions at later times, the third path (6), 
which is triggered by the $^3\rm{H}(\alpha,\gamma)^7\rm{Li}(n,\gamma)^8\rm{Li}(\alpha,n)^{11}\rm{B}$ chain reactions, becomes equally important. See the text
for more details.
\label{fig4}}
\end{figure}
\end{center}

\section{Quest for Nuclear Physics and Astrophysics}

We discussed the r-process nucleosynthesis
in a single set of $(M, L_{\nu})$ of the neutrino-driven wind in sect. 2.
In realistic supernova events, however, the neutrino luminosity varies slowly 
as a function of time during Kelvin-Helmholtz cooling phase $\sim 10$ sec, as
observed in SN1987A.  
We therefore expand the parameter space in this section in order to include a number of
$(M, L_{\nu})$-grids in the reasonable range $1.2 M_{\odot} \leq M
\leq 2.0 M_{\odot}$ and 
$10^{50} {\rm ergs/s} \leq L_{\nu} \leq 5 \times 10^{52}$ ergs/s.  
Calculated results of $\tau_{\rm dyn}$ and $S$ are displayed
in Fig. 5~\cite{otsuki00}.  
Shown also are two zones for which the r-process nucleosynthesis might occur
so that the second abundance peak ($A \approx 130$) 
and the third abundance peak ($A \approx 195$) emerge 
from a theoretical calculation~\cite{hoffman97}.
In this figure, we observe that the neutrino-driven 
winds from massive and compact proto-neutron stars may produce successfully
the r-process elements.

In order to compare the nucleosynthesis result with the
observed $r$-process abundance, the calculated 
yields are integrated over the mass-weighted time history,
according to the time evolution of the neutrino luminosity
\begin{equation}
L_\nu = L_{\nu 0} \exp (-t/\tau_\nu),
\end{equation}
where we set $L_{\nu 0} = 5 \times 10^{52}$~ergs~s$^{-1}$~\cite{woosley94} 
and $\tau_\nu =1.0$~s so that the total integrated neutrino energy
becomes $3 \times 10^{53}$~ergs to fit the observed result in SN1987A.
Assuming that the trajectories can be described at each time 
by steady flow~\cite{otsuki00} corresponding to the grids shown in Fig. 5,
the mass-weighted r-process yields are given~\cite{wanajo01} by
\begin{equation}
Y_i = \frac{1}{M_{\rm ej}} \int Y_i (t) \dot M (t) dt, 
\end{equation}
\begin{equation}
M_{\rm ej} = \int \dot M (t) dt,
\end{equation}
where $Y_i$ is the number fraction of i-th nuclide, $\dot M (t)$ is 
the mass-loss rate, and $M_{\rm ej}$ is 
the total mass ejected in the neutrino-driven winds
from the neutron star.  

The calculated result is shown in Fig. 6.  Total mass ejection is 
$M_{\rm ej} = 1.5 \times 10^{-3} M_{\odot}$, and total mass of the r-process elements 
$M_{\rm r-process} = 1.0 \times 10^{-4} M_{\odot}$.  
These values are compared with those of the benchmark study
by Woosley et al.~\cite{woosley94},
which gives $M_{\rm ej} = 3 \times 10^{-2} M_{\odot}$ and 
$M_{\rm r-process} = 1.0 \times 10^{-4} M_{\odot}$.  
Their calculation makes larger total mass ejection because their highest entropy is 
three times bigger and the expansion time scale is about one order 
of magnitude larger than those in the present wind model.  
Unfortunately, their result turns out to be incomplete~\cite{meyer95}
for the lack of neutrino-nucleus interactions during the nucleosynthesis,
as discussed in the previous section.  We included these interactions fully in the present
calculation.  Having the same $M_{\rm r-process}$, 
we can explain the absolute mass of the observed r-process abundance
in the solar system, taking account of the frequency of finding 
SNeII in the Milky Way over $10^{10}$ yr.

Our result in Fig. 6 reproduces fairly well the observed 
second abundance peak (A $\approx$ 130), the hill of rare-earth elements (A $\approx$ 165), 
and the third peak (A $\approx$ 195).  
However, there are several defects, too.  
The biggest defect is that the abundance around A $\approx 100 \sim 120$
is overproduced by, at most, factor 50.  
Although this factor is much smaller than the overproduction 
by two or more orders of magnitudes in the previous 
calculations~\cite{woosley94,meyer92}, 
this is a common defect in the hot-bubble scenario.
This overproduction is related apparently to the hydrodynamics of the
expanding neutrino-driven winds.  These nuclei are produced in the
$\alpha$-rich freeze out.
One of the hints to resolve this difficulty may be the bypass effects 
due to the neutrino interactions with very neutron-rich nuclei~\cite{mathews01}.
Further studies are needed to suppress this unwelcome peak.  

The second defect is a shift of the third abundance peak around A $\approx$ 195
by a couple of mass units.  This is also a common feature found in the previous
r-process calculations~\cite{woosley94,otsuki00,wanajo01,meyer92}.  
These elements are the beta-decay products of extremely neutron-rich
unstable nuclei on the neutron magic N = 126.  
Peak position depends on the timing of freezeout of the r-process and
subsequent beta-delayed neutron emission.
Therefore, a particular combination of environmental evolution of 
neutron-number density, $N_n$,
and temperature, $T_9$, as well as the expansion dynamic time scale,
$\tau_{dyn}$, might match the freeze out so that it results in the
right position of the abundance peak.
\begin{center}
\begin{figure}
\hspace{2cm}\includegraphics[width=85mm,angle=-90]{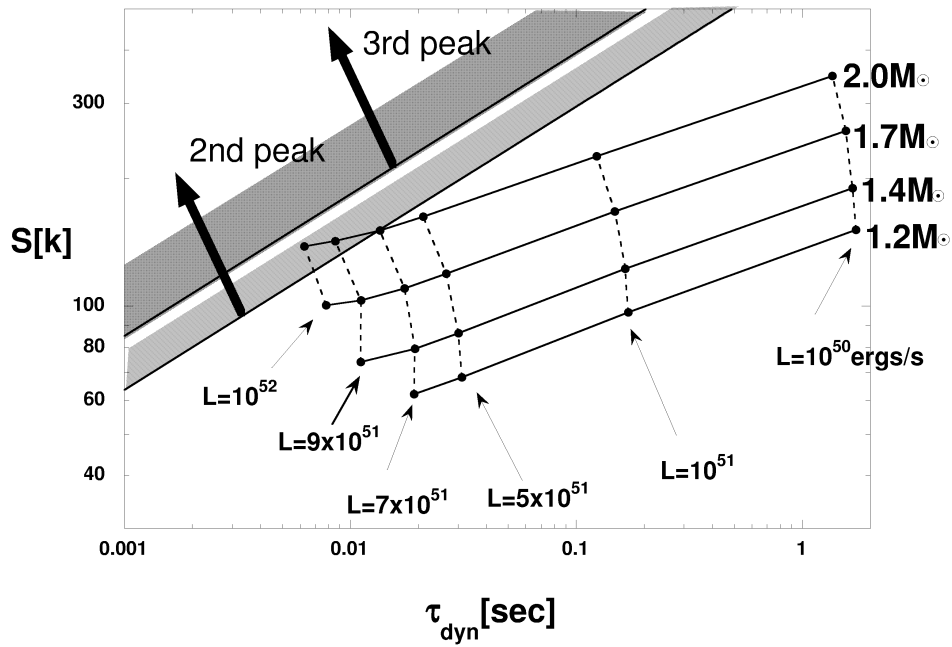}
\caption{
Relation between entropy per baryon, S/k, and dynamic time scale,
$\tau_{dyn}$, for various combinations of the neutron star mass,
$1.2M_{\odot} \leq M \leq 2.0M_{\odot}$, and the 
neutrino luminosity, $10^{50} \leq L_{\nu} \leq 10^{52}$ ergs/s,
taken from Otsuki et al.~\cite{otsuki00}.
Solid and broken lines connect the same mass and luminosity.
Two zones indicated by shadows  satisfy the approximate conditions, for
$Y_e=0.4$, on which the successful r-process occurs~\cite{hoffman97}
to make the second abundance peak around $A=130$ (lower) and 
the third abundance peak around $A=195$ (upper).
\label{fig5}}
\end{figure}
\end{center}

The third defect is the deficiency of abundance 
right above or below the peak elements,
i.e. at A $\approx 140 \sim 150$ and $\approx 180 \sim 190$.
These deficiencies seem related to yet unseen effects of deformation
or strucure changes of unstable nuclei surrounding the neutron magic 
numbers N = 82 and 126.
Further extensive theoretical studies and observational challenge
to determine the masses, lives, and beta Q-values of these nuclei are highly desired.

The fourth defect is the underproduction of actinoid elements, 
Th-U-Pu (A = 230 $\sim$ 240), 
by more than one order of magnitude.
The observed high abundance level of these nuclei (see Fig.~6)
might suggest an existence of a new magic number 
around N = $150 \sim 160$.
Xenon $^{129}$Xe$_{75}$ and platinum $^{195}$Pt$_{117}$ are the typical 
r-process elements on the second and third abundance peaks,
which are the decay products from extremely neutron-rich unstable 
nuclei with neutron magic numbers N = 82 and 126, respectively.
From these observations we estimate that the waiting point nucleus 
is located by shifting $\Delta$N $\approx$ 7 or 9 units from the peak element.
Applying the same shift $\Delta$N $\approx$ 7 or 9 to $^{232}$Th$_{142}$ and
$^{238}$U$_{146}$, we could assume a new magic number 
around N = $150 \sim 160$ which may lead to 
the fourth abundance peak at A = 230 $\sim$ 240. 
Actually, in the very light nuclear systems, a new magic number N = 16
was found~\cite{ozawa00} in careful experimental studies of the neutron 
separation energies and interaction cross sections of extremely neutron-rich nuclei.
Since these possibilities were not taken into account
in the present and previous calculations, 
the deficiency of actinoids might be improved by modernizing nuclear mass formula
including such effects.

Another possibility is to make actinoid elements in neutron star mergers
or the mergers of the neutron star and black hole binaries which have
extremely small lepton fraction, $Y_e \le 0.2$~\cite{freiburghaus99}.
However, these processes do not virtually produce any 
intermediate-mass nuclei including iron,
which contradicts with the fact that the observed iron abundance
is at least proportional to the r-process elements and actinoid elements
over the entire history of Galactic evolution $\sim 10^{10}$ yr.
Different possibility has recently been proposed in~\cite{sumiyoshi01}.
Inner mass shells of the prompt supernova explosions have very small 
lepton fraction, $Y_e \le 0.16$, and thus some actinoid elements 
are produced here.

Since we discuss only material ejected from the proto-neutron star
behind the shock, it does not make any serious problem 
to see the underproduction in mass region A $\le$ 90.
Most of these intermediate-mass nuclei are ejected from the
exploded outer shells in supernovae.
\begin{center}
\begin{figure}[htb]
\hspace{3cm}\includegraphics[width=100mm]{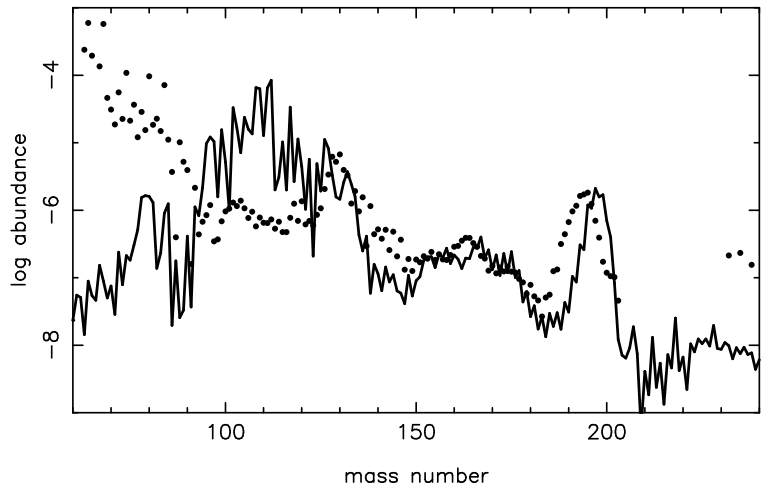}
\caption{
Final mass-weighted r-process abundance (solid line) as a function of atomic mass number
$A$ compared with the solar system r-process abundance (filled circles)~\cite{kappeler89}.
The neutrino-driven wind model used is for
$L_{\nu} = 10^{52}$ ergs/s and $M = 2 M_{\odot}$.
The solar system r-process abundance is shown in arbitrary unit. 
\label{abundance}}
\end{figure}
\end{center}

\section{Quest for Astronomy}

The solar system abundance is thought to consist of many different nucleosynthesis 
products in various astrophysical sites.  For this reason it is not straightforward to 
compare the "pure" process to be studied theoretically with those observed experimentally
in the solar system. 
However, recent detections~\cite{sneden96} of extremely enhanced r-process 
abundances in several metal-deficient stars have indicated that 
they describe surprisingly very similar pattern
to the solar system r-process abundance which we call a "universality".
We here proved theoretically that the neutrino-driven winds from 
massive and compact proto-neutron stars 
can explain this "universality".

Let us refer to massive neutron stars.  
Although a large core mass $1.7 M_{\odot} \le M$ is sometimes claimed
from observational viewpoints, several established EOSs
for neutron star matter are known to stabilize these massive cores.  
Wanajo et al.~\cite{wanajo01} have recently proposed a formation mechanism of 
massive neutron star in the delayed explosion mechanism.  
Since the r-process nucleosynthesis is very
sensitive to the core mass of proto-neutron stars~\cite{otsuki00,wanajo01,sumiyoshi00},
it is desirable to probe the mass dependence of their yields observationally.
Incidentally, a large dispersion in heavy element abundances of halo stars 
has recently been observed~\cite{mcwilliam95,ryan96}.
Ishimaru and Wanajo~\cite{ishimaru99} have shown in their galactic chemical
evolution model that if r-process nucleosynthesis occurs in either
massive supernovae $\geq 30 M_{\odot}$ or small mass supernovae $8-10
M_{\odot}$, the observed
large dispersion can be well understood.
In addition, SN1994W and SN1997D are presumed to be due to 25
$M_{\odot}-40 M_{\odot}$ massive progenitors because of a very low $^{56}$Ni 
abundance in the ejecta~\cite{SN1,SN2}.
These massive supernovae are known to have massive iron cores $\geq 1.8
M_{\odot}$ and leave massive remnants~\cite{SN2}.

It is more critical for the r-process nucleosynthesis whether the remnant is
a neutron star or a black hole.
The recent discovery~\cite{israelian99} of a strong overabundance
of $\alpha$-elements in the companion in the massive
black hole binary, GRO J1655-40, suggests clearly that
the explosion was a gravitational type core-collapse, i.e. SNeII.
The atmosphere of the star orbiting around black hole or neutron star
is likely to be polluted and contaminated by the nucleosynthesis products
in supernova event associated to the formation of these compact objects.  
Therefore, if a future spectroscopic analysis of this companion star
could detect the r-process elements and if the
mass of the black hole were determined observationally, 
it would show clear evidence for the fact that the r-process 
could occur in SNeII which leave remnant black holes.

It has recently been suggested theoretically that, 
even should the core mass be $M \sim 1.4 M_{\odot}$ which is a typical cold
neutron star mass, 
taking proper boundary conditions of the wind at the shock r $\sim$ 10,000 km 
can improve crucial conditions for the r-process nucleosynthesis.  
Extensive studies are underway~\cite{terasawa01new}.

The r-process elements in the first generation stars might also make a
strong impact on cosmology.
The actinoids like U-Th-Pu are produced only in the r-process,
and for their long lives, which are comparable to the cosmic expansion age, 
these elements have ever been used as cosmochronometers.
Cosmological models infer that the cosmic expansion time scale is 
$10^{10}$ y $\le t_U \le 1.5 \times 10^{10}$ y, but there are several other
observations which suggest controversially longer Galactic age 
than $t_U$.  Major source of the uncertainty arises from the 
uncertain cosmological distance scale.
In order to construct a reliable nuclear cosmochronometer, which has
a great advantage being free from the distance scale, 
one has to establish that the atmosphere of 
extremely metal-deficient stars has 
for the most part originated from the gas of
a single supernova event which produced enough amount of r-process elements
in the early Galaxy.
Extensive theoretical studies~\cite{otsuki01new} of the r-process
have suggested that the "universality" may hold only for elements
having $Z = 56 \sim 80$, while it seems slightly broken for the actinoid elements.
They have suggested also that the Th/U chronometer is almost free from
this breaking universality.
These theoretical suggestions have recently been confirmed by the
detections of Th along with other r-process elements in several new 
metal-poor halo giants which were observed by using the High Dispersion Spectrograph 
equipped with SUBARU Telescope, SUBARU/HDS~\cite{honda01}. 
Further collaboration 
among nuclear physics, astrophysics, astronomy, and even cosmology
is highly desirable.

This work was supported in part
by Japan Society for Promotion of Science, and by the Grant-in Aid for
Scientific Research (1064236, 10044103, 11127220, and 12047233) of the
Ministry of Education, Science, Sports and Culture of Japan, and DoE
Nuclear Theory grant DE-FG02-95-ER40394.



\end{document}